\documentclass{pnastwo}

\advance\voffset -.5in % Minus dimension will raise the printed page on the
\usepackage{amssymb,amsfonts,amsmath}
\usepackage{graphicx}
\usepackage{url}
\usepackage{hyperref}
\newcommand{\rem}[1]{}

\contributor{Submitted to Proceedings
of the National Academy of Sciences of the United States of America, ET's inaugural article}
\copyrightyear{2008}
\issuedate{Issue Date}
\volume{Volume}
\issuenumber{Issue Number}
\begin{document}

%%%%%%%%%%%%%%%%%%%%%%%%%%%%%%
%% For titles, only capitalize the first letter
\title{Static and dynamic friction in sliding colloidal monolayers}

%% Enter authors via the \author command.
%% Use \affil to define affiliations.
%% (Leave no spaces between author name and \affil command)
%% \author{<author name>
%% \affil{<number>}{<Institution>}} One number for each institution.
%% The same number should be used for authors that
%% are affiliated with the same institution, after the first time
%% only the number is needed, ie, \affil{number}{text}, \affil{number}{}
%% Then, before last author ...
%% \and
%% \author{<author name>
%% \affil{<number>}{}}
%% For example, assuming Garcia and Sonnery are both affiliated with
%% Universidad de Murcia:
%% \author{Roberta Graff\affil{1}{University of Cambridge, Cambridge,
%% United Kingdom},
%% Javier de Ruiz Garcia\affil{2}{Universidad de Murcia, Bioquimica y Biologia
%% Molecular, Murcia, Spain}, \and Franklin Sonnery\affil{2}{}}

\author{
Andrea Vanossi\affil{1}{CNR-IOM Democritos National Simulation
    Center, Via Bonomea 265, 34136 Trieste, Italy}\affil{2}{International
    School for Advanced Studies (SISSA), Via Bonomea 265, 34136 Trieste,
    Italy},
Nicola Manini\affil{3}{Dipartimento di Fisica, Universit\`a degli
    Studi di Milano, Via Celoria 16, 20133 Milano,
    Italy}\affil{2}{}\affil{1}{}
\and
Erio Tosatti\affil{2}{}\affil{1}{}\affil{4}{International Center
    for Theoretical Physics (ICTP), Strada Costiera 11, 34104 Trieste,
    Italy}
}

\contributor{Submitted to Proceedings of the National Academy of Sciences
of the United States of America}

%% The \maketitle command is necessary to build the title page.
\maketitle

%%%%%%%%%%%%%%%%%%%%%%%%%%%%%%%%%%%%%%%%%%%%%%%%%%%%%%%%%%%%%%%%
\begin{article}

\begin{abstract}
%MAX 250 WORDS, PNAS instructions say it should be on page 2 of the manuscript!
%
In a pioneer experiment, Bohlein {\it et al.} realized the controlled
sliding of two-dimensional colloidal crystals over laser-generated periodic
or quasi-periodic potentials.
Here we present realistic simulations and arguments which besides
reproducing the main experimentally observed features, give a first
theoretical demonstration of the potential impact of colloid sliding in
nanotribology.
The free motion of solitons and antisolitons in the sliding of hard
incommensurate crystals is contrasted with the soliton-antisoliton pair
nucleation at the large static friction threshold $F_{\rm s}$ when the two
lattices are commensurate and pinned.
The frictional work directly extracted from particles' velocities can be
analysed as a function of classic tribological parameters, including speed,
spacing and amplitude of the periodic potential (representing respectively
the mismatch of the sliding interface, and the corrugation, or ``load'').
These and other features suggestive of further experiments and insights
promote colloid sliding to a novel friction study instrument.
\end{abstract}

%% When adding keywords, separate each term with a straight line: |
% NICK: ADD MORE KEYWORDS?

%% Optional for entering abbreviations, separate the abbreviation from
%% its definition with a comma, separate each pair with a semicolon:
%% for example:
%% \abbreviations{SAM, self-assembled monolayer; OTS,
%% octadecyltrichlorosilane}
% \abbreviations{}

%% The first letter of the article should be drop cap: \dropcap{}
%\dropcap{I}n this article we study the evolution of ''almost-sharp'' fronts

\dropcap{T}he intimate understanding of sliding friction, a central player
in the physics and technology of an enormous variety of systems, from
nanotribology to mesoscale and macroscale sliding
\cite{VanossiArXiv11,Urbakh04}, is historically hampered by a number of
difficulties.
One of them is the practical inaccessibility of the buried interface
between the moving bodies -- with few exceptions, we can only hypothesize
about its nature and behavior during sliding.
Another is the general impossibility to fully control the detailed nature,
morphology, and geometric parameters of the sliders; thus for example, even
perfectly periodic, defect-free contacting surfaces have essentially only
been accessible theoretically.
If we knew and, on top of that, if we could control the properties and the
relative asperity parameters of the sliders, our physical understanding
could greatly increase, also disclosing possibilities to tune friction in
nano and mesoscopic systems and devices.
As Bohlein {\it et al.}~\cite{Bohlein12} showed, two dimensional (2D)
colloid crystalline monolayers can be forced by the flow of their embedding
fluid to slide against a laser-generated static potential
mimicking the interface ``corrugation'' potential in ordinary
sliding friction.
The external pushing force, the interparticle interactions, and especially
the corrugation potential are all under control, the latter ranging from
weak to strong, and from periodic, to quasi-periodic
\cite{Mikhael08,Mikhael10}, in principle to more complex types too.
Contrary to established techniques in meso and nanosize sliding friction
(Atomic Force Microscope, Surface Force Apparatus, Quartz Crystal
Microbalance) \cite{Carpick97}, which address the tribological response in
terms of averaged physical quantities (overall static and kinetic friction,
mean velocities, slip lengths and slip times, etc.), in colloid sliding
every individual particle can in principle be followed in real time,
stealing a privilege hitherto restricted to the ideal world of
molecular-dynamics (MD) simulations
\cite{Vanossi12,Reichhardt02,Reichhardt11}.

Materializing concepts long-anticipated theoretically
\cite{Braunbook,Braun97}, the colloid sliding data showed how the sliding
of a flat crystalline lattice on a perfectly periodic substrate takes place
through the motion of soliton or antisoliton superstructures (also known in
one dimension -- 1D -- as kinks or antikinks) -- positive or negative
density modulations that reflect the misfit dislocations of the two
lattices that are incommensurate in their mutual registry.
While forming regular static Moir\'e superstructure patterns when at rest,
solitons constitute the actual mobile entities during depinning and
sliding, and are essential for ``superlubricity'' \cite{Erdemir07} -- i.e.,
zero static friction -- of hard incommensurate sliders.
When solitons are absent at rest owing to commensurability of the two
sliders (or are present but pinned in soft incommensurability), the colloids
and the periodic potential are initially stuck together.
Only after the static friction force $F_{\rm s}$ is overcome, solitons
appear (or depin if they already exist but are pinned) unlocking the
colloids away from the corrugation potential, so that sliding can take
place.

Our aim here is to understand and demonstrate, based on molecular dynamics
(MD) sliding simulations, how the great colloid visibility and
controllability can be put to direct use in a tribological context.
The full phase diagram versus colloid density and sliding force is explored
first of all, highlighting a large asymmetry between solitons and
antisolitons, and a strong evolution
from commensurate to
incommensurate caused by sliding.
We then extract and predict the frictional work as a function of mean
velocity and corrugation amplitude, loosely mimicking ``load'' -- the same
variables of classic macroscopic friction laws.
We also discuss new local phenomena underlying depinning, including the
edge-originated spawning of incommensurate antisolitons and the
bulk-originated nucleation/separation of soliton-antisoliton pairs in the
commensurate case, as well as global analogies to driven Josephson
junctions, charge-density waves, and the sliding of adsorbate islands on
crystal surfaces
\cite{Ustinov88,Gruener88,Krim88,Krim91,Tomassone97,Bruschi02,Bruschi06}.

\section{Modeling and simulations}

\begin{figure}
\centerline{
\includegraphics[angle=0,width=0.44\textwidth,clip=]{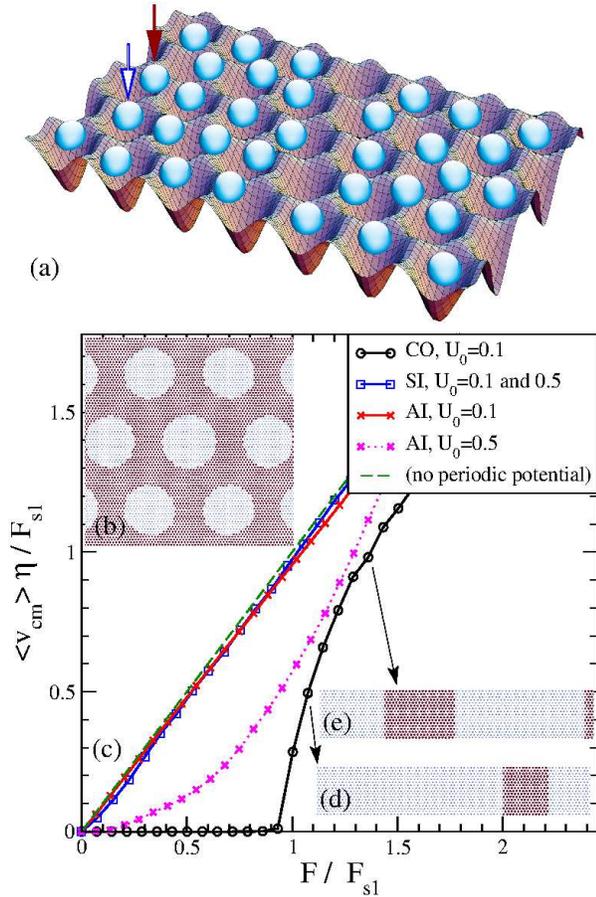}
}
\caption{ \label{mobility:fig} % (Color online)
(a) A sketch of the model for the colloid particles interacting with a
  periodic potential $W$.
(b) The static initial configuration at $\rho= 0.95$.
Three families of antisoliton lines (darker areas) cross at 120 degrees.
(c) Velocity-force characteristics for various colloid densities,
with a lattice-potential corrugation commensurability ratio $\rho= a_{\rm
las}/a_{\rm coll}=1.0$ (CO), $1.05$ (SI), and $0.95$ (AI).
The CO case always displays static friction.
For weak corrugation ($U_0=0.1$), $F_{\rm s}=0$ in both AI and SI
incommensurate cases.
At larger corrugation ($U_0=0.5$) a major asymmetry appears between the AI
and SI configurations: only the AI case exhibits a finite depinning
threshold with static friction.
(d,e) Snapshots of the central region of the initially commensurate colloid
during motion, illustrating sliding-generated solitons, whose density
increases as $F$ is increased.
In all snapshots, colloids located at repulsive spots of the corrugation
potential [defined by $W({\mathbf r})>-U_0/2$, e.g.\ the colloid pointed at
  by the red filled arrow in panel (a)] are drawn as dark red spots, while
colloids nearer to potential minima [$W({\mathbf r})\leq -U_0/2$, e.g.\ the
  colloid pointed at by the blue empty arrow in panel (a)] are light blue.
}
\end{figure}

The driven colloids are modelled as charged point particles undergoing
overdamped 2D planar dynamics under an external force $F$, parallel to the
plane, applied to each colloid.
While the fluid is not described explicitly, $F$ is to be interpreted as
$\eta v_d$ where $\eta$ and $v_d$ are the effective fluid viscosity and
velocity.
Particles repel each other with a screened Coulomb interparticle repulsion,
$V(r_{ij}) = Q/r_{ij} \, \exp(-r_{ij}/\lambda_D)$ with $\lambda_D$
substantially smaller than the mean distance between particles.
Colloids are immersed in a Gaussian-shaped overall confining potential
$G(|{\mathbf r}|) = -A_c \exp(-r^2/\sigma^2)$ -- the large radius $\sigma$
representing the laser spot size -- and in a triangular-lattice periodic
potential $W({\mathbf r}) = -(2 U_0/9) \large[ 3/2 +2\cos(2\pi r_x/a_{\rm
    las})\cos(2\pi r_y/(\sqrt{3}a_{\rm las})) +\cos(4 \pi r_y/(\sqrt{3}
  a_{\rm las})) \large]$ representing the interface ``corrugation'',
see Fig.~\ref{mobility:fig}a.
Finally, in addition to the external force, a Stokes viscous force
$-\eta\, \mathbf v_i$ acts on each particle $i =1,...,N$,
% NICK: added following referee's question 4
and accounts for the dissipation of the colloids kinetic energy into the
thermal bath.
We typically simulate $N \simeq 30,000$ -- a particle number much smaller
than in experiment, but sufficient to extract reliable physical results.\footnote{
To reduce simulation sizes and times, our particles form an island near the
center of the Gaussian potential, which is the region experimentally
visualized. Particles outside this region, whose role is less relevant, are
omitted.
Moreover, thermal effects (although straightforward to introduce in
simulation), have not been accessed in colloid experiments.
After verifying that the main features are not washed out at 300~K 
(see supporting information), here we will, for the sake of clarity, only
present results obtained with a dissipative $T=0$ Langevin dynamics.
} % end of footnote
In the absence of corrugation ($U_0=0$), colloids form a 2D crystalline
island at rest.
The 2D density of the triangular 2D lattice is fixed by $N$ and by the
balance of the confining energy $G$ and the 2-body repulsion energy.
We set this balance so that the average colloid lattice spacing $a_{\rm
  coll}$ (before
submittal to the corrugation potential $W$) is unity.\footnote{
The spacing of the fully relaxed colloid configuration varies smoothly from
$a \simeq 0.984$ at the sample center to $a\simeq 1.05$ at the side, with
an average density equal to that of a triangular crystal of spacing $a_{\rm
  coll}=1$.
} % end of footnote
We then realize a variety of mismatched ratios $\rho= a_{\rm las}/a_{\rm
  coll}$ by changing the corrugation period $a_{\rm las}$.
In the following, we focus on three representative cases, namely:
underdense, $\rho= 0.95$ (antisoliton-incommensurate -- AI; the starting
state at rest is shown in Fig.~\ref{mobility:fig}b); ideally dense, $\rho=
1.0$ (nearly commensurate -- CO, which becomes exactly commensurate after
turning on $W$); overdense, $\rho = 1.05$ (soliton incommensurate -- SI)
\footnote{
Similar models were studied in the past with a view to understand two
dimensional Frenkel-Kontorova models and adsorbate monolayers physics
\cite{Lomdahl86,Srolovitz86,Gornostyrev99,Braun01b}}.
In order to simulate experiment, and also to prevent solitons from leaving
the finite-size sample, our external force $F$ is ramped
in time in small well-spaced steps,
so that its overall value slowly alternates
in sign, forth and back with a long time period.
Full simulation details are given in the supporting information.

\section{Results}

\begin{figure}
\centerline{
\includegraphics[angle=0,width=0.42\textwidth,clip=]{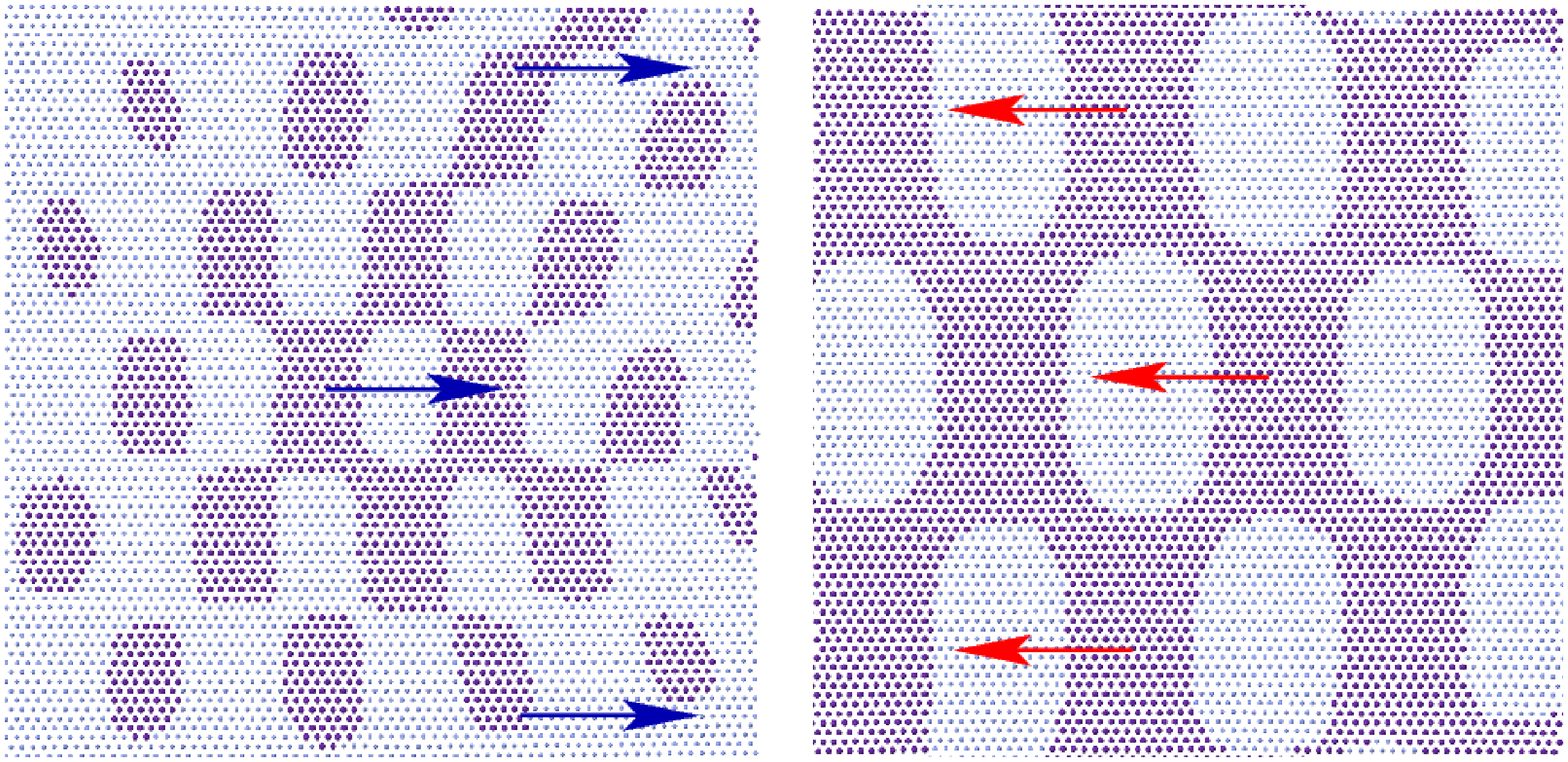}
}
\caption{ \label{patterns:fig} % (Color online)
Depinned, moving particles (darker) of a 2D colloid in a periodic potential
under the action of a rightward force $F$.
(a) Rightward propagating solitons of overdense colloids  ($\rho=1.05$, SI);
(b) Leftward propagating antisolitons of underdense colloid ($\rho=0.95$, AI).
}
\end{figure}
%%%%%%%%%%%%%%%%%%%%%%%%%%%%%%%%%%%%%%%%%%%%%%%%%%%%%%%%%%%%%

Figure~\ref{mobility:fig}c displays the mean speed $\langle v_{\rm
  cm}\rangle$ of the central portion of the colloid system as a function of
the driving force $F$.
Fully reproducing experiment \cite{Bohlein12}, the simulated force-velocity
characteristics of Fig.~\ref{mobility:fig}c show a large static
friction force threshold in the $\rho\simeq 1$ CO case, where the
colloid and corrugation lattices are pinned together. Static friction is lost in
case of incommensurability and moderate corrugation, where preformed mobile
solitons or antisolitons are present.
The snapshots of Fig.~\ref{patterns:fig} illustrate the patterns of
solitons/antisolitons sliding in opposite directions under the same driving
force $F>0$.
For a weak external force and a $\sim 5\%$ lattice mismatch, the static
friction drops essentially to zero, and a nearly free viscous sliding is
realized, reflecting a situation of ``superlubricity''
\cite{Peyrard83,Dienwiebel04,Filippov08}.
However, under the same conditions, not all incommensurate geometries are
superlubric.
Whereas for weak corrugation the overall colloid mobility $\langle v_{\rm
  cm}\rangle/F$ is remarkably constant for both incommensurate densities,
we find in fact that by increasing the corrugation amplitude $U_0$ the
mobility of the AI configuration drops to zero at small force, and pinning
with static friction reemerges despite incommensurability.
By contrast, SI configurations remain superlubric up to much larger $U_0$.

\subsection{The Aubry transition}
Borrowing results of the 1D Frenkel Kontorova (FK) model \cite{Braunbook},
the single-soliton width
$d\simeq  g^{1/2} a_{\rm las}$
%Braunbook pag. 6
where $g= \xi a_{\rm las}^2 k / U_0$
%Braunbook pag. 3
[here $k = V''(a_{\rm coll})$ and $\xi$ is a constant of order unity]
is large for a hard layer on a weak corrugation, and small for a soft layer
on a strong corrugation.
Between these two extremes, the 1D incommensurate FK model crosses the
so-called Aubry transition \cite{Peyrard83} where superlubricity is lost,
and pinning sets in with static friction despite incommensurability.
Even in the present 2D case it is qualitatively expected that all
incommensurate colloids, both underdense ($\rho\lnsim 1$) and overdense
($\rho\gnsim 1$) will undergo an Aubry-like superlubric-to-pinned
transition for increasing corrugation.

\begin{figure}
\centerline{
\includegraphics[angle=0,width=0.48\textwidth,clip=]{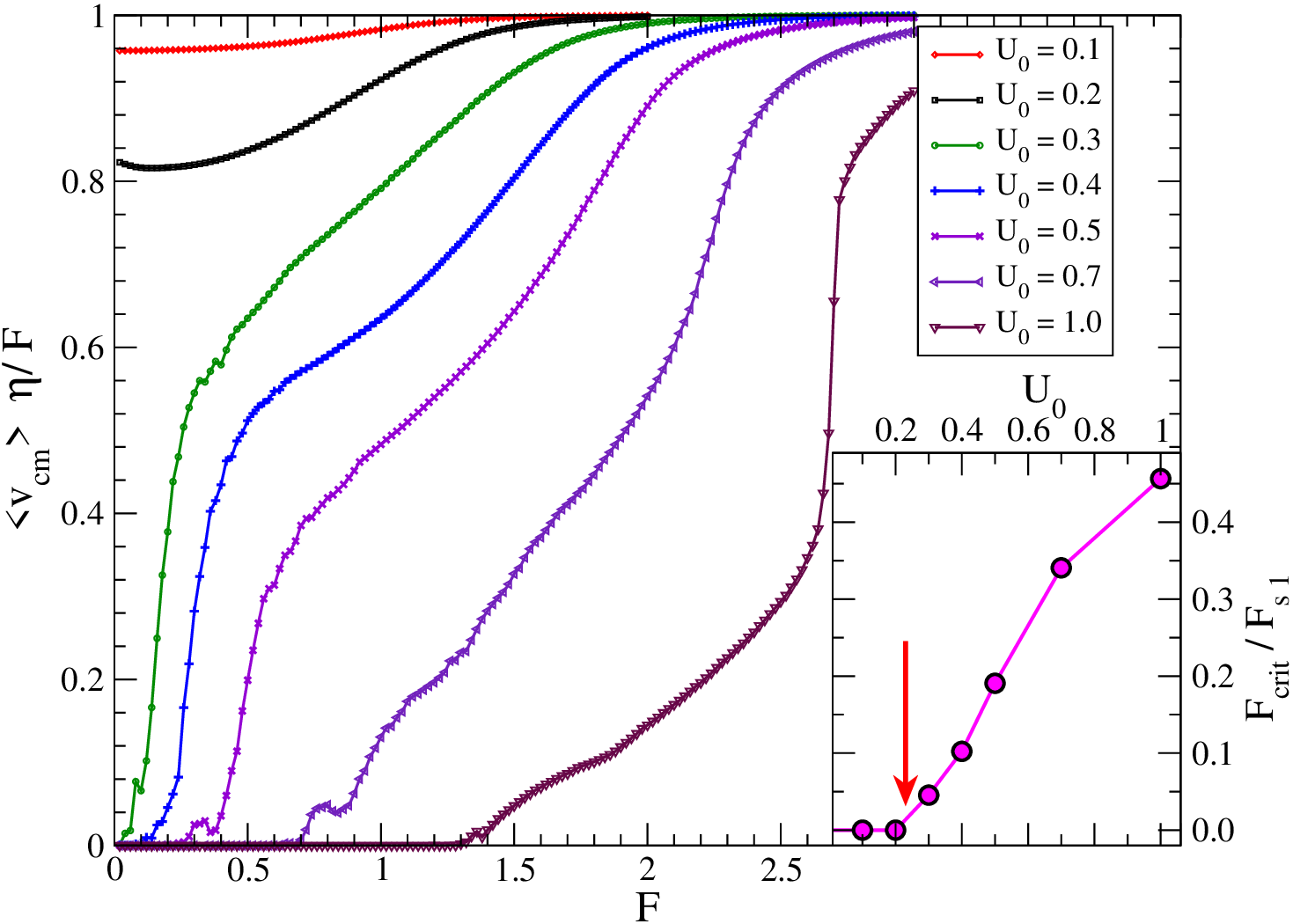}
}
\caption{ \label{bulk_anti_transition:fig} % (Color online)
2D Aubry transition for antisolitons at $\rho = 0.95$, in an infinite-size
colloid system.
Main panel: colloid mobility as a function of the applied force, for
increasing corrugation amplitude $U_0$. Note the appearance of pinning
with static friction just above $U_0$= 0.2.
Inset: static friction (depinning) force $F_{\rm s}$, normalized to the
single-colloid force barrier $F_{s\,1}$, as a function of $U_0$,
with an arrow indicating the critical Aubry corrugation.
}
\end{figure}

This expectation is confirmed in our 2D model colloid system.
Figure~\ref{bulk_anti_transition:fig} (obtained by independent simulations
of the infinite-size system with periodic boundary conditions) shows the
Aubry-like pinning transition crossed by an AI ($\rho=0.95$) underdense
colloid at a critical corrugation, here $U_{0}^{\rm crit}\simeq 0.2 - 0.3$.
The threshold Aubry corrugation depends upon $\rho$, and is much larger for
overdense SI than for underdense AI colloids.

\subsection{Soliton-antisoliton asymmetry}

This strong asymmetry of static friction -- and of all other properties --
between overdense ($\rho \gnsim 1$) and underdense ($\rho \lnsim 1$)
colloids can be rationalized, in the limit of strong corrugation
$g \ll 1$,
in terms of the large physical difference between solitons, defects
formed by lines of lattice interstitials, and antisolitons, lines of
vacancies.
This asymmetry remains even for weak corrugation ($g \gg 1$),
when solitons/antisolitons involve
relative displacements far smaller than those of proper interstitials or vacancies.
A small variation $\delta$ in the inter-colloid separation $a$ is
sufficient to produce a large relative variation of the effective spring
constant, i.e.\ the interaction curvature
\begin{equation}
\frac{V''(a\pm \delta)}{V''(a)} \simeq
%\frac{V'(a\pm \delta)}{V'(a)} \simeq
\frac{V(a\pm \delta)}{V(a)} \simeq
\exp(-\delta/\lambda_D)
.
\end{equation}
For a realistic $\lambda_D=0.03\,a_{\rm coll}$, this highly nonlinear and
asymmetric relation, implies a huge $460\%$ increase whenever two colloids
are approached by $5\%$ of their average separation, but only a $82\%$
reduction for a $5\%$ increased separation.
This asymmetry is held responsible for the much weaker propensity of
solitons to become pinned and to localize compared to antisolitons.

\subsection{The sliding state}

Under sliding, the shapes and geometries of solitons/antisolitons and their
motion are of most immediate interest, as they are directly comparable with
experiment.
Figure~\ref{patterns:fig} shows the large-scale checkerboard structure of
solitons/antisolitons of the sliding colloid lattice.
They move with a speed $v$ much larger than the average lattice speed
$\langle v_{\rm cm}\rangle$, because $v/\langle v_{\rm cm}\rangle \sim
\rho/(\rho-1)$ by particle conservation.
The moving structure is a distortion of the original triangular
soliton/antisoliton pattern (Fig.~\ref{mobility:fig}b) induced by the
circular shape of the confining potential, and by the directional sliding.
With increasing $F$, the soliton %shapes
arrangements
elongate into a stripe-like
pattern perpendicular to the driving direction.
Comparison with experimental pictures is quite realistic,
especially when focusing (as
done in experiment) on the central sample region, far from boundaries.

In the AI superlubric colloid $\rho\lnsim 1$, preformed antisolitons fly
(leftward) across the colloid lattice antiparallel to the (rightward)
force.
They are eventually absorbed at the left edge boundary, while new ones
spawn at the right edge boundary to replace them, sustaining a steady-state
mobility.
In the SI superlubric colloid $\rho \gnsim 1$ conversely, preformed
solitons fly rightward, parallel to the force.
Solitons, unlike antisolitons, are not automatically spawned at the
boundary, owing to the decreasing density.
Instead, an antisoliton/soliton pairs must nucleate first, near the
boundary, and this is possible only if the force overcomes the nucleation
barrier.
Below this threshold, we observe that a steady DC external force eventually
sweeps out all the preformed solitons transforming the colloid to an
artificially pinned, immobile CO state.

\begin{figure}
\centerline{
\includegraphics[angle=0,width=0.35\textwidth,clip=]{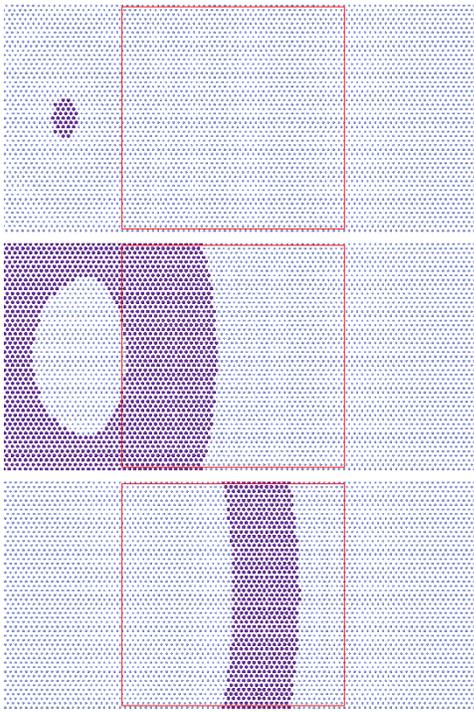}
}
\caption{ \label{nucleation} % (Color online)
Three successive snapshots of the initial depinning instants of the
commensurate ($\rho=1$) configuration for the $F\simeq F_{s\,1}$
simulation, see Fig.~\ref{mobility:fig}.
The horizontally extended window visualizes the the nucleation and
separation of a soliton-antisoliton pair, left of the central observation
region (square).
Pair nucleation constitutes the depinning mechanism of all commensurate
sliders.
}
\end{figure}

Finally, the pinned CO colloid $\rho \gtrsim 1$ only moves after static
friction is overcome.
As illustrated in Fig.~\ref{nucleation}, motion starts off here by
nucleation of soliton-antisoliton pairs inside the bulk -- here close to
the left edge because the central region tends to be slightly overdense.
The antisolitons flow leftwards and are absorbed by the left edge, becoming
undetectable to the optically monitored central part of the colloid, where
only solitons transit, as seen in experiment.
This type of commensurate nucleation has been described in considerable detail
in literature, including finite-temperature effects
\cite{Braun97,Reguzzoni10}.
We note here that in the pinned CO colloid the soliton or antisoliton
density, initially zero, actually increases with increasing sliding
velocity (see, e.g., Fig.~\ref{mobility:fig}d,e), as opposed to frankly
incommensurate cases, where it is nearly constant.

\subsection{Phase-diagram evolution with sliding}

\begin{figure}
\begin{tabular}{c}
\includegraphics[angle=0,width=0.42\textwidth,clip=]{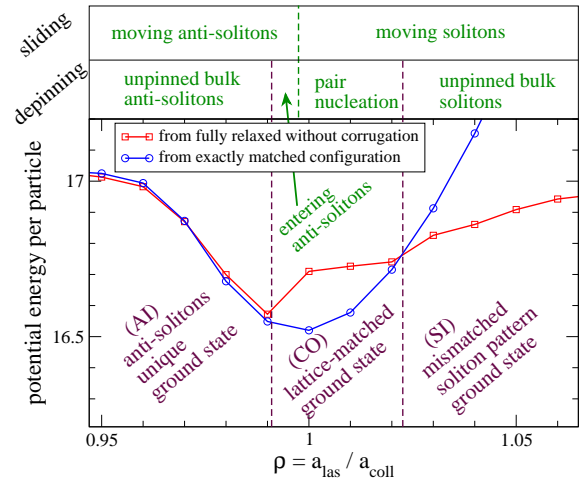}\\
\end{tabular}
\caption{ \label{static:fig} % (Color online)
Effective ``phase diagram'' of the finite colloid crystal as a function of
the lattice spacing mismatch $\rho$, for $U_0=0.1$.
The two potential energy curves (colloid-colloid repulsion plus $W$
interaction) characterize the static phases:
(red squares) relaxation started from the slightly inhomogeneous
configuration produced by a previous relaxation for $U_0=0$;
(blue circles) relaxation started with a fully matched lattice of spacing
$a_{\rm las}$.
Two AI (underdense) and SI (overdense) phases surround a commensurate phase
(CO).
The depinning mechanisms and the soliton structures sustaining sliding
are illustrated at top of figure for the different regions.
}
\end{figure}

Much can be learned about the habit of sliding colloids from their behavior
and their structural phase diagram, first at rest and then under sliding.
With $\rho\simeq 1$, close to commensurate but not exactly commensurate,
the colloid monolayer can realize in the periodic potential two alternative
static arrangements which are local minima of the overall free energy: a
fully lattice-matched CO state, or a weakly incommensurate state
characterized by a sparse soliton (AI or SI) superstructure, with a
density fluctuating around the local value prescribed by the $G - V$
balance.
Comparing the potential energy of these two states as a function of $\rho$,
the static phase diagram contains, as sketched in Fig.~\ref{static:fig}, a
fully commensurate extended CO region separated from the AI and SI regions
by commensurate-incommensurate transitions, well known in adsorbed surface
layers \cite{Coppersmith81,Bak82,Patrykiejew99,Mangold03}.
The CO region is wider on the SI side ($\rho>1$) than the AI side
($\rho<1$), another manifestation of the SI-AI asymmetry discussed above.
The CO range naturally widens or shrinks when the corrugation amplitude
$U_0$ is increased or decreased.

Under an external force $F$, sliding effectively tilts the balance between
the two static phases
(loosely speaking, for of course under sliding  the physical significance of a ``phase'' is
not the same as at rest) sliding populates the former CO phase with
solitons/antisolitons, turning it effectively into SI or AI.
In the running state, the colloid average density increases or decreases
from 1 to a value closer to the nominal $\rho$ of the colloid at $U_0 =
0$.
This explains why in a quasi-commensurate configuration with $\rho \gnsim 1$
such as that shown by Bohlein {\it et al.}~\cite{Bohlein12}, solitons (and not, e.g.,
soliton-antisoliton pairs) sweep the colloid
upon depinning, as also seen in
Fig.~\ref{mobility:fig}d,e.

It is curious to note here the different fate of solitons in the slightly
overdense CO and in the SI phases.
In the CO phase they do not exist at rest, but they appear after depinning
and under sliding.
In the SI phase they exist at rest, but they could
be swept out under DC sliding,
when a weak external force can
turn the SI colloid into effectively CO.
We never saw this sweepout phenomenon on the AI side.

\section{Frictional analysis}

We turn now to frictional work, a quantity of crucial importance for the
tribological significance of colloid sliding.
We can write the overall power balance as the scalar product of the
instantaneous velocity $\mathbf v_i$ of each colloid $i$ by the net force
acting on it, $\eta \left( \mathbf v_d - \mathbf v_i \right)$, including
both the bare external force $\mathbf F=\eta \mathbf v_d $ and the viscous
drag $- \eta \mathbf v_i$.
This product vanishes instantaneously at any time when either
colloids are stuck ($\mathbf v_i = \mathbf 0$) or else when the corrugation
potential is absent, so that $\mathbf v_i \equiv \mathbf v_d$.
After averaging over a very long trajectory, the balance reads
\begin{eqnarray}\nonumber
P_{\rm tot}&=&\sum_i
\eta \langle \left( \mathbf v_d - \mathbf v_i \right) \cdot \mathbf v_i \rangle
\\\nonumber
&=&\left( N \mathbf F \cdot \langle \mathbf v_{\rm cm} \rangle
  - \eta \langle |\mathbf v_{\rm cm}|^2 \rangle\right)
- \eta  \sum_i \langle | \mathbf u_i|^2\rangle
\\\label{power:eq}
&=& P_{\rm frict} - P_{\rm kin}
\,,
\end{eqnarray}
where $ \mathbf u_i = \mathbf v_i - \mathbf v_{\rm cm}$.
Under steady-state sliding conditions where $P_{\rm tot}=0$, the effective
friction power $P_{\rm frict}$ is exactly balanced by an internal
kinetic energy excess rate.
Per colloid particle, $P_{\rm frict}$ is
\begin{equation}\label{power1:eq}
p_{\rm frict} =
\frac{P_{\rm frict}}N \simeq
\mathbf F \cdot \langle \mathbf v_{\rm cm} \rangle
 - \eta |\langle \mathbf v_{\rm cm}\rangle|^2
,
\end{equation}
where small center of mass fluctuations are neglected, by assuming $\langle
{\mathbf v_{\rm cm}}^2\rangle \simeq |\langle \mathbf v_{\rm cm}\rangle|^2$.

\begin{figure}
\centerline{
\includegraphics[angle=0,width=0.48\textwidth,clip=]{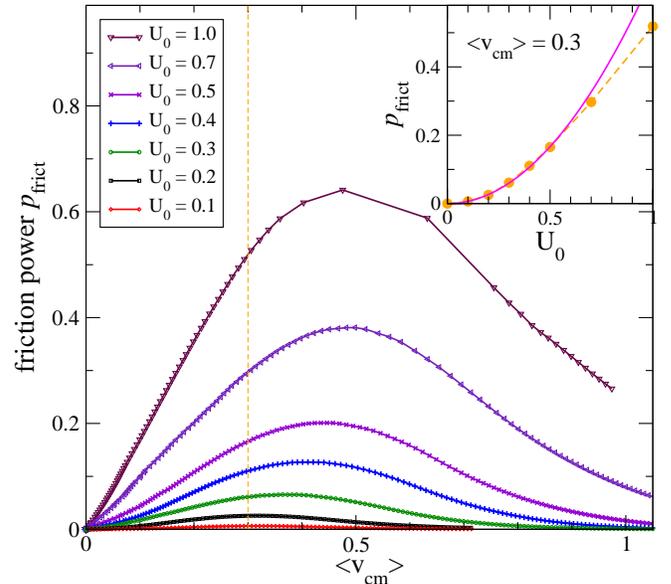}
}
\caption{ \label{bulk_anti_friction:fig} % (Color online)
Sliding friction $p_{\rm frict}$ per particle as a function of speed
$\langle v_{\rm cm}\rangle$ for an underdense AI colloid $\rho = 0.95$ for
increasing corrugation amplitude $U_0$.
Inset: $U_0$ dependence of friction for speed $\langle v_{\rm cm} \rangle =
0.3$, showing the quadratic rise for weak corrugation behavior, followed by
a roughly linear growth.
}
\end{figure}

Figure~\ref{bulk_anti_friction:fig} shows $p_{\rm frict}$ (briefly referred
to as ``friction'' in the following) for the special case of the AI
underdense phase as extracted as a function of $\langle \mathbf v_{\rm
  cm}\rangle$ through a ``bulk'' simulation (with periodic boundary
conditions as in Fig.~\ref{bulk_anti_transition:fig}).
The main features found are (i) a linear rise at low CM speed; (ii) a decline at large
speed; (iii) a maximum at some intermediate corrugation-dependent speed.
We moreover observe that (iv) the dissipated power increases (not
unexpectedly) with corrugation; and (v) the corresponding frictional
maximum simultaneously shifts to larger speed.

The qualitative interpretation of these results is relatively
straightforward, and yet revealing.
(i, iv) At low sliding velocities the motion of solitons/antisolitons
involves the viscous motion of individual particles with a velocity
distribution whose spread toward higher values rises proportionally to the
sliding speed and inversely proportional to their spatial width.
As shown, e.g.\ within the 1D FK model \cite{Braunbook}, but also in the
present simulations, the width $d$ of solitons/antisolitons increases
roughly as $d \sim a_{\rm las} \sqrt g$ with the dimensionless
interparticle interaction strength $g\propto a_{\rm las}^2 V''(a_{\rm
  coll}) / U_0$ measured relative to the periodic corrugation amplitude.
The decrease in width with increasing corrugation $U_0$ requires an
increasing instantaneous speed of individual particles in the
soliton/antisoliton, yielding an increasing viscous friction, and a
decreasing overall mobility as observed.
(ii, iii, v) At high sliding velocities, the colloid relaxation time
exceeds the soliton/antisoliton transit time across the Peierls-Nabarro
barrier \cite{Braunbook} so that their spatial structure is gradually
washed out by the sliding motion.
The critical speed where the smoothening behavior takes over, roughly
corresponding to maximal friction, increases as corrugation increases,
corresponding to narrower solitons/antisolitons that are harder to wash
out.
The increase of friction with corrugation strength $U_0$, plotted in the
inset for a chosen speed, is found to be quadratic at weak corrugation,
gradually turning to linear for larger values.
Linear response theory naturally accounts for the quadratic increase, a
behavior first discussed by Cieplak {\it et al.}\ \cite{Cieplak94} and
observed in quartz crystal microbalance experiments \cite{Coffey05}.

Demonstrated for a specific AI case with antisolitons, the above results
appear in fact of general validity for infinitely extended sliders of
controlled colloid density, and apply equally well although with great
quantitative asymmetry to SI with solitons, once their larger widths and
greater mobilities and weaker Peierls-Nabarro barriers are taken into
account.

\section{Summary and Conclusions}

In this study we have presented initial simulation results and theory that
strongly vouch in favor of sliding of colloid layers on laser-originated
corrugations as a promising tool for future tribological advances.
The motion of solitons and antisolitons known from experiment is reproduced
and understood, unraveling the subtle depinning mechanisms at play.
The presence of Aubry transitions is pointed out for future verification,
along with a strong asymmetry between underdense and overdense
incommensurate layers.
Of direct tribological interest, we anticipate the behavior of friction
with corrugation (mimicking ``load'') and with sliding velocity, with
results which, while of course generally very different from the classic
laws of macroscopic friction, are highly relevant to friction at nano and
mesoscopic scales.
Our approach moreover indicates a strong complementarity between theory plus
simulation, and experiment, an aspect which we intend to pursue further.

There are many lines of future research that this study implicitly
suggests.
One line will be to pursue the analogy of the sliding over a periodic potential
with other systems such as driven Josephson junctions \cite{Tinkham96}, and
sliding charge-density waves \cite{Gruener88}.
Time-dependent nonlinear phenomena such as the Shapiro steps
\cite{Gruener88,Tinkham96} should become accessible to colloid sliding too.
A second line is to include non-periodic complications to the corrugation
potential, including the quasicrystal geometry such as that recently
realized \cite{Bohlein12PRL} and beyond that, random, or pseudo-random
corrugations to be realized in the future.
A third line involves the investigation of the lubricant speed quantization
phenomena, characterized so far only theoretically
\cite{Vanossi06,Cesaratto07,Vanossi07PRL,Castelli09}.

A further very important development will be to address colloidal friction
in larger, mesoscopic or macroscopic size systems, whose phenomenology is
accessible so far only by a few, very ingenious, but very limited, methods
\cite{Carpick97,Krim91,Tomassone97, Drummond01,Rubinstein04,Rubinstein06}.
A major scope in that case will be to realize and study stick-slip friction
and aging phenomena, at the heart of realistic physical and technological
tribology.

%% == end of paper:

%% Optional Materials and Methods Section
%% The Materials and Methods section header will be added automatically.

%% Enter any subheads and the Materials and Methods text below.
%\begin{materials}
% Materials text
%\end{materials}

%% Optional Appendix or Appendices
%% \appendix Appendix text...
%% or, for appendix with title, use square brackets:
%% \appendix[Appendix Title]

\begin{acknowledgments}
\label{AcknSect}

This work is partly funded by
the Italian Research Council (CNR) via Eurocores FANAS/AFRI,
by the Italian Ministry of University and Research through PRIN projects
20087NX9Y7 and 2008y2p573, and by the Swiss National Science Foundation
Sinergia CRSII2\_136287.

\end{acknowledgments}

% Sample bibliography item in PNAS format:
%% \bibitem{in-text reference} comma-separated author names up to 5,
%% for more than 5 authors use first author last name et al. (year published)
%% article title  {\it Journal Name} volume #: start page-end page.
%% ie,
% \bibitem{Neuhaus} Neuhaus J-M, Sitcher L, Meins F, Jr, Boller T (1991)
% A short C-terminal sequence is necessary and sufficient for the
% targeting of chitinases to the plant vacuole.
% {\it Proc Natl Acad Sci USA} 88:10362-10366.
%\bibliographystyle{unsrt}
%\bibliography{biblio}

\end{article}

\end{document}